\documentclass[12pt,apjl]{emulateapj}
\usepackage{natbib}

\newcommand{\kepler}{\textit{Kepler}\ }
\usepackage{multirow}

\shortauthors{Morris et al.}
\shorttitle{Secondary Eclipse \& Planet-Induced Gravity Darkening of HAT-P-7}

\begin{document}
\title{Kepler's Optical Secondary Eclipse of HAT-P-7b and Probable Detection of Planet-Induced Stellar Gravity Darkening}

\author{Brett M. Morris\altaffilmark{1,2}}
\author{Avi M. Mandell\altaffilmark{2}}
\author{Drake Deming\altaffilmark{1,2}}
\altaffiltext{1}{Department of Astronomy, University of Maryland, College Park, MD 20742, USA}
\altaffiltext{2}{Goddard Center for Astrobiology, NASA's Goddard Space Flight Center, Greenbelt MD 20771, USA}
\slugcomment{Accepted: ApJL, Jan 18, 2013}

\begin{abstract}
We present observations spanning 355 orbital phases of HAT-P-7 observed by \kepler from May 2009 to March 2011 (Q1-9). We find a shallower secondary eclipse depth than initially announced, consistent with a low optical albedo and detection of nearly exclusively thermal emission, without a reflected light component.  We find an approximately 10 ppm perturbation to the average transit light curve near phase -0.02 that we attribute to a temperature decrease on the surface of the star, phased to the orbit of the planet.  This cooler spot is consistent with planet-induced gravity darkening, slightly lagging the sub-planet position due to the finite response time of the stellar atmosphere. The brightness temperature of HAT-P-7b in the \kepler bandpass is $T_B = 2733 \pm 21$ K and the amplitude of the deviation in stellar surface temperature due to gravity darkening is approximately $-0.18$ K. The detection of the spot is not statistically unequivocal due its small amplitude, though additional Kepler observations should be able to verify the astrophysical nature of the anomaly.
\end{abstract}

\keywords{Planets and satellites: individual: HAT-P-7b --- Planets and satellites: fundamental parameters --- Stars: Individual: HAT-P-7 --- Stars: atmospheres --- Eclipses}

\section{Introduction}
Since launching in 2009, the \kepler Space Telescope has provided unparalleled photometry of more than 160,000 stars, with the primary goal of detecting transiting extrasolar planets.  By providing continuous short-cadence photometry across a single field of view, \kepler has succeeded in detecting thousands of new candidate transiting planets (see \citealp{borucki2011} or \citealp{batalha2012} for a recent review), but it has also provided extremely high-precision photometric time series for three extrasolar planets that were known prior to launch: TrES-2b, HAT-P-11b and the hottest of the bright \kepler planets, HAT-P-7b.  These targets, due to the brightness of the parent stars and their well-constrained orbital parameters, represent an opportunity to search for transit-curve features that are unobservable for any other objects at present, and which can inform us about the characteristics of the planet-star system. In this Letter we report on our analysis of HAT-P-7b, the brighter and hotter of the three objects.

HAT-P-7b was originally discovered by \citet{pal2008} in the HATNet project. This transiting hot Jupiter with mass $1.7 ~ M_J$ orbits an F6 star with an orbital semi-major axis only four times greater than the radius of its host star. HAT-P-7b is one of the hottest known transiting exoplanets in the \kepler field, and atmospheric models by \citet{madhusudhan2010}, \citet{spiegel2010} and \citet{christiansen2009}, hereafter denoted C10, suggest strong evidence for thermal inversion.

As a known bright exoplanet target, HAT-P-7 was given high priority by \kepler and has been observed using a one-minute cadence nearly continuously since May 2009. Using only the first 10 days of observations, \citet{borucki2009} were able to detect the secondary eclipse at optical wavelengths, finding an eclipse depth of $130 \pm 11$ parts per million (ppm) as well as orbital phase variations. \citet{welsh2010} determined that these phase variations were dominated by perturbations due to ellipsoidal variations in the shape of the host star caused by the gravity of the planet, resulting in brightening of the system on either side of the secondary eclipse with an amplitude on the order of $\sim$60 ppm (Jackson et al.\ 2012, hereafter denoted J12).

Each of these studies was limited by the amount of data available at the time -- the most recent study by J12 analyzed data up through Quarter 2. With the recent release of newly re-calibrated data spanning Q1-Q9, an updated analysis of the full set of transits and eclipses (326 transits and 355 eclipses in all, after selection criteria are applied) is warranted. In this work we focus on analyzing both the variability of the eclipse depths, as well as the characteristics of the mean transit and eclipse curves. These measurements allow us to place improved constraints on the presence of cloud variability and planetary albedo, and to search for additional effects of the planet on the host star. In \S\ref{sec:observations} we describe our observational data set, in \S\ref{sec:analysis} we describe our transit fitting methods and analysis methodology, and in \S\ref{sec:discussion} and \S\ref{sec:summary} we discuss and summarize our results. 

\section{Observations} \label{sec:observations}
Short-cadence ($\sim$60 s) photometry of HAT-P-7 was recorded by \kepler from 2009 May 13 -- 2011 June 26 or Quarters 1-9. More than one million photometric measurements produced by the \kepler data analysis pipeline were retrieved. We examined fluxes calculated from both SAP (Simple Aperture Photometry) and PDCSAP (Presearch Data Conditioned Simple Aperture Photometry). The conditioning process, a standard component of the \kepler data pipeline, seeks to remove systematic effects introduced by the spacecraft while retaining interesting astronomical signals \citep{jenkins2010}. The results we show here were calculated with the PDCSAP fluxes, since we expect these data to contain less uncorrected systematic effects, and thus to yield more accurate results. We carried out the analysis with both types of flux and found that the results were consistent with each other, despite the additional corrections imposed on the PDCSAP data set.

Each successive orbital phase of observations of the planet was considered for inclusion in our analysis individually. We rejected orbital phases that contained uncorrected systematic effects after the \kepler pipeline removal and any observation that did not record a complete time series on phase --0.2 to 0.2 (near the transit) or 0.3 to 0.7 (near the eclipse), where the phase is defined on 0-1, with zero centered on mid-transit. 
\begin{figure}
\epsscale{1}
\plotone{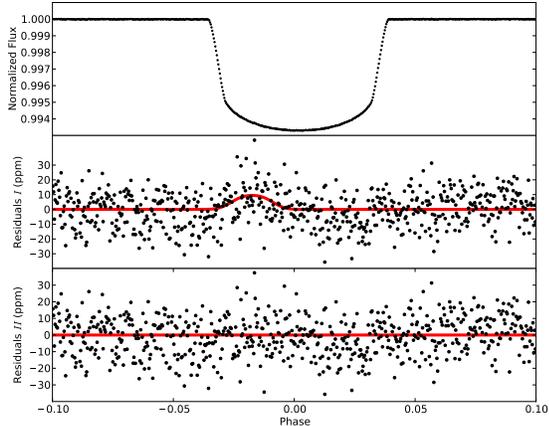}
\caption{\textit{Upper}: Composite transit light curve of HAT-P-7b from 326 transits. The $>$400,000 photometric measurements in the range of phase from -0.1 to 0.1 are averaged in 1-minute (334 photometric measurement) bins. Error-bars are omitted for clarity.
\textit{Middle}: Residuals after removing the best-fit \citet{mandel2002} light curve model using Levenburg-Marquardt least squares. The curve is the best-fit model light curve for the anomaly in the \citet{mandel2002} model caused by the planetary transit across the gravity-darkened spot on the surface of the star.
\textit{Lower}: Residuals after removing the minimum-$\chi^2$ gravity-darkened spot model (circles) and the expectation after removal of the \citet{mandel2002} and gravity-darkened spot models (line at 0 ppm). The standard deviation of these final residuals is 11.6 ppm.}
\label{phaseFoldedTransit} 
\end{figure}

\section{Analysis} \label{sec:analysis}
\subsection{Primary Transit}
\begin{table*}
\caption{HAT-P-7 System Parameters}
\begin{center}
\begin{tabular}{l l r}\\
Parameter & Definition & \kepler Photometry\\
\hline\hline\\
$P$ (days) & Period & $2.204737 \pm 0.000017$\\ \\
$T_c$ (BJD TDB) & Epoch & $ 2454954.357463 \pm 0.000005$ \\ \\
$i$ (deg) & Inclination & $83.111 \pm 0.030$ \\ \\
\multirow{2}{*}{$R_p/R_*$}  & \multirow{2}{*}{Ratio of planetary} & \multirow{2}{*}{$0.07759 \pm 0.00003 $} \\ \\
&  radius to stellar radius & \\ \\
$\gamma_1$ & Limb-darkening, linear& $0.3525 \pm 0.0066 $ \\ \\
$\gamma_2$ & Limb-darkening, quadratic & $0.168 \pm 0.010$ \\ \\
\multirow{2}{*}{$a/R_*$} & \multirow{2}{*}{Ratio of semimajor axis}& \multirow{2}{*}{$4.1502 \pm 0.0039 $} \\ \\
& to stellar radius & \\ \\
$D$ (ppm) & Secondary eclipse depth & $69.1 \pm 3.8$\\ \\ 
$T_B$ (K) & Brightness temperature & $ 2733 \pm 21$\\ \\ 
& Phase at mid-eclipse &  $0.500051 \pm 0.000006$\\ \\ 
\hline 
\end{tabular}
\end{center}
{Ephemerides (i.e., $P$ and $T_c$) derived from Levenburg-Marquardt least-squares fits to each transit, and orbital parameters derived by the Levenberg-Marquardt least-squares fit to the composite transit light curve with one-minute binning. The uncertainties of the parameters were calculated using the prayer-bead method. The linear and quadratic limb-darkening coefficients ($\gamma_1$ and $\gamma_2$) are defined as in the \citet{mandel2002} formalism.} 
\label{tab:orbParam}
\end{table*}
Using the orbital parameters determined by C10, we calculated an initial estimate for the mid-transit time for each transit. Model light curves of the form derived by \citet{mandel2002} were then fit to each transit over orbital phase --0.2 to 0.2 using Levenburg-Marquardt least-squares. We fit for the phase angle of mid-transit, ratio of planetary to stellar radii ($R_p/R_*$), ratio of the semi-major axis of the planet to the radius of the star ($a/R_*$), inclination ($i$), zero-point flux and linear and quadratic limb darkening coefficients ($\gamma_1$ and $\gamma_2$); the newly-derived mid-transit times were then used to determine an improved orbital period (see Table \ref{tab:orbParam}). We also searched for periodicities in the deviations of the mid-transit times from our best-fit ephemeris with the Lomb-Scargle algorithm, but found no evidence for significant transit timing variations. 

The new ephemerides were used to construct a phase-folded transit light curve; we then fit a transit model to the composite light curve with Levenberg-Marquardt least-squares, including a binning of the model to match the sampling of the data \citep{kipping2010}. Uncertainties calculated using the ``prayer-bead'' method \citep{gillon2007}. The orbital parameters derived from this fit are listed in Table \ref{tab:orbParam}. Our linear limb-darkening coefficient is consistent with model atmosphere predictions by \citet{claret2011}, but our required quadratic limb-darkening coefficient is moderately different than their predicted values.

\subsection{Stellar Gravity Darkening}
The residuals of the composite transit light curve of HAT-P-7b (see Figure \ref{phaseFoldedTransit}) show an apparent brightening of the system just prior to mid-transit. After fitting for the transit model, we found $\chi^2 \approx 720$.  This is greater than the expected value for the 642 photometric measurements with seven fitting parameters ($\chi^2_{ideal} = 635$), giving a p-value of 1\%. Thus the anomaly in the transit light curve is of marginal statistical significance to verify with certainty that it has an astrophysical origin. However, we will hypothesize here that it is real and test candidate astrophysical phenomena that can cause a similar light curve anomaly. 

Transient stellar phenomena such as magnetically-driven star spots can be ruled out as the cause of the anomaly, since star spots only persist for time scales much shorter than this set of observations, which spanned two years. To ensure that the anomaly is not a few star-spot crossings that became diluted in the average, we fit model transit light curves to four subsets of the entire observation, and the anomaly was persistent in each subset. We therefore require a periodic stellar dimming associated with the planet's orbital period. \citet{barnes2009} suggests that gravity-darkening of rapidly rotating stars can perturb transit light curves; however this effect is only significant for stars that rotate much more rapidly than HAT-P-7, which has $v \sin{i} = 3.8$ km/s \citep{pal2008} and an expected rotation velocity $v \sim 15$ km/s \citep{winn2009}. Furthermore, gravity darkening due to rotation affects the light curve on all times between the second and third contact, whereas the anomaly that we detect is highly-localized in phase angle. One possible explanation for such a localized anomaly is planet-induced gravity darkening, in which the tidal distortion of the star due to the planet causes a decrease in the surface gravity near the sub-planet region, which would result in a corresponding decrease in stellar surface brightness (J12). Here, we construct a simple model to characterize the plausibility of gravity darkening as an explanation for the anomaly in the residuals.

\begin{figure}
\epsscale{1}
\plotone{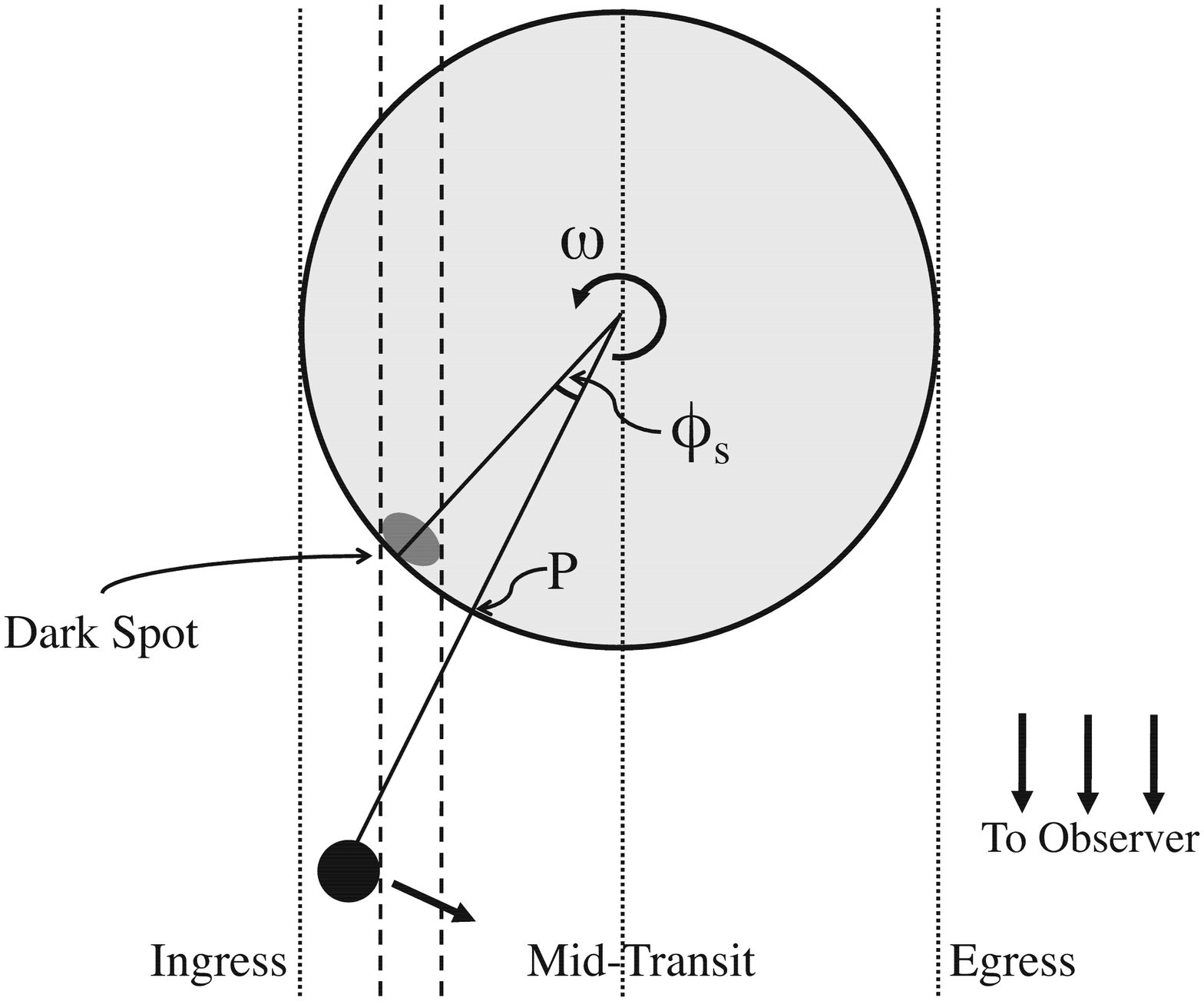}
\caption{The system geometry is depicted above, with the planet represented by the small dark circle, and the star represented by the large grey circle, with a gravity-darkened spot on its surface. The transit of the planet across the dark spot occurs after ingress and before mid-transit, so long as the phase lag between the sub-planet point P and the centroid of the spot, $\phi_s$, is small and nonzero. The diagram depicts the beginning of the transit across the dark spot, viewed at an angle such that the orbital plane is in the plane of the page. Note that
the plane containing the planet and the observer is tilted by $90-i = 6.9$ degrees to the plane of the page. The schematic is not drawn to scale.}
\label{fig:diagram} 
\end{figure}

The dark spot is assumed to follow a Gaussian intensity profile with its centroid located in the orbital plane of the planet, displaced from the sub-planet point on the stellar surface by some phase lag, $\phi_s$, as illustrated in Figure \ref{fig:diagram}. We calculate the brightness of the system as the planet transits and the dark spot traverses the surface of the star. The total flux as a function of orbital phase is then compared to a similar model with no dark spot to demonstrate the difference between our model and the expected transit light curve. The difference predicts
the amplitude, phase lag and width ($\sigma_s$) of the anomaly in the transit light curve residuals. 

We fit our spot model to the residuals of the transit light curve (Figure \ref{phaseFoldedTransit}) to determine the amplitude and phase lag of the darkened spot, then we convert the amplitude of the anomaly in ppm to Kelvins using a blackbody model for the star. The model which minimizes the $\chi^2$ computed between that model and our composite transit light curve is adopted as the ``best-fit'' model. We described the emission spectrum of HAT-P-7 with that of a blackbody with an effective temperature of $T_* = 6350$ K (as measured by \citealp{pal2008}), since conventional stellar atmospheric models do not simulate the differences between distinct atmospheres with temperature differences on the order of tenths of Kelvin. We found a ``best-fit'' spot model with phase lag of the Gaussian centroid from the sub-planet point on the stellar surface  $\phi_s =Ð0.056$ [phase units], with amplitude --106 ppm and radial spread $\sigma_s = 0.037$ [phase units]. We found that the most consistent temperature difference between the darkened spot and the mean stellar surface, calculated from this amplitude, is $T_{spot} - T_* =  -0.183$ K, consistent with the anomaly of ``a few 0.1 K'' predicted by J12.

Using our derived orbital inclination and $a/R_*$, we find that as the planet transits the spot, the location of the sub-observer point (the projection of the transit onto the star) is 29.6 degrees above the line of sight to the center of the star.  If we assume that the path of the center of the spot has the same inclination as the planetary orbit (6.9 degrees from the line-of-sight to the star), the fact that the planet transits the spot at all means that the spot must have a half-width of $\ge22.7$ degrees in the direction orthogonal to the plane of the planet's orbit.  We found $\sigma_s$, the 1-sigma width of the spot deduced from the fit to the transit across the spot in the direction of the orbit of the planet, to be $\sigma_s = 0.032$, or 11.5 degrees. Assuming that the dark spot is real, it is evidently more extended in the direction orthogonal to the plane of the planet's orbit.

\subsection{Secondary Eclipse}
As mentioned above, several previous studies have measured the mean depth for the secondary eclipse of HAT-P-7b, but these results are all significantly different from each other.  \citet{borucki2009} found an eclipse depth of $130 \pm 11$ ppm, while \citet{welsh2010} and J12 found secondary eclipse depths of $85.8$ ppm and $61 \pm 3$ ppm, respectively.  Each study used a different number of observations, as well as different methods to compensate for the planetary phase variations and the stellar ellipsoidal variations. With the goal of removing model uncertainties, we masked the eclipse and fit model phase curves to the light curve to account for the significant phase variation effects, Doppler beaming, planetary reflection and ellipsoidal variation curves \citep{faigler2011}, from phase 0.3 to 0.7. We then divided by this baseline before fitting to the eclipse. We hereafter chose to present results from the PDCSAP fluxes, which are consistent with the results from the SAP fluxes.	

\begin{figure} 
\epsscale{1}
\plotone{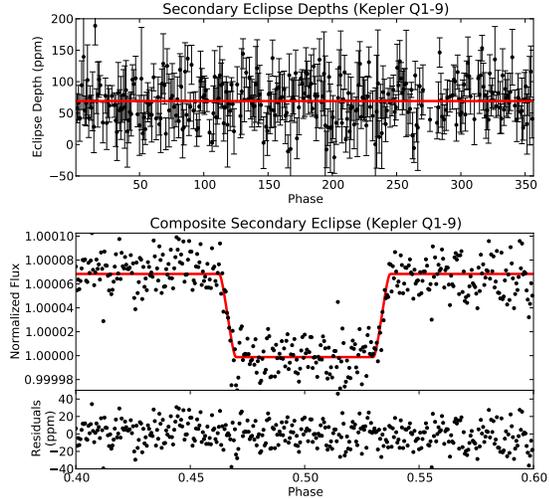}
\caption{\textit{Upper:} Eclipse depth measurements for each phase analyzed, with best-fit depth = $69.1 \pm 3.8$ ppm (horizontal line). This depth is consistent with predominantly thermal emission, with nearly no reflected light component. 
\textit{Lower:} Phase-folded eclipse light curve for Kepler Quarters 1-9. The photometric measurements are averaged in $\sim1.5$ minute bins (circles). Error-bars are omitted for clarity. The curve represents the \citet{mandel2002} model light curve for a secondary eclipse of a uniformly luminous occulted planet.}
 \label{eclipseDepth}
\end{figure}

We fit a \citet{mandel2002} model eclipse light curve to each eclipse individually using Levenberg-Marquardt least-squares. The orbital parameters found by the previous fit to the composite transit light curve (see Figure \ref{phaseFoldedTransit}) were used to fix the duration of the eclipse and the mid-eclipse times; uncertainties for each depth measurement were estimated by the prayer-bead method. We then derived the mean eclipse depth by computing a least-squares fit to the eclipse depth over all phases included in our analysis of Quarters 1-9, yielding a depth of $69.1 \pm 3.8$ ppm (see Figure \ref{eclipseDepth}). The mid-eclipse time derived from the fit to the composite secondary eclipse light curve occurs virtually exactly at the expected value (phase = 0.5), after taking light travel time across the orbit into account.  This value is significantly lower than the eclipse depth found by \citet{borucki2009}, due to both the longer observational baseline and the inclusion of ellipsoidal variations in our model; fitting only the data analyzed by \citet{borucki2009}, we find an eclipse depth of $105 \pm 27$ ppm.

The individual eclipse depths exhibit scatter of 32 ppm over the full data set, assuming a constant expected value. This scatter is higher than can be accounted for using the \kepler-provided uncertainties (designated ``SAP\_ERR'' and ``PDCSAP\_ERR''). In order to ensure that the scatter we measured does not include intrinsic variability in the eclipse depth, we repeated the analysis to search for an eclipse at phase=0.3, where no eclipse is expected. Our analysis is consistent with no eclipse at phase=0.3 with 32 ppm rms scatter, which confirms that the scatter between measurements is indeed non-astrophysical. This complete analysis was repeated using both the SAP and PDCSAP fluxes and similar measurements were obtained for the eclipse depth and scatter. 

Scaling the geometric albedo result from C10 with our smaller measurement of the eclipse depth, we find a geometric albedo of $\lesssim 0.03$. This low optical albedo of HAT-P-7b is consistent with observations of other highly irradiated hot Jupiters, such as TrES-2b  with albedo $\sim0.01$ (see \citealp{barclay2012} and \citealp{kipping2011a}) and HD 209458 \citep{rowe2008}. Theoretical predictions also call for low albedos for hot Jupiters, such as the atmospheric models by C10.

To compute the brightness temperature of the planet, a Kurucz model star with temperature $T_*=6350$ K is adopted for HAT-P-7. If we assume that all of the flux from the atmosphere of the planet is thermal emission, the ratio of the blackbody flux of the planet to the Kurucz flux of the star integrated over the \textit{Kepler} bandpass gives the secondary eclipse depth. These considerations yield a brightness temperature in the \kepler bandpass of $T_B =2733 \pm 21$ K, where these uncertainties propagate from the rms scatter of the eclipse depth (see Figure \ref{eclipseDepth}).

\section{Discussion}\label{sec:discussion}
To further test the hypothesis that the anomaly is caused by a gravity-darkened spot on the stellar surface, we compare the physical characteristics of the dark spot to our knowledge of stellar atmospheres. The dark spot is offset in phase from the sub-planet point on the stellar surface by phase=--0.059. In other words, about 3.1 hours elapse at the sub-observer point on the stellar surface between the passing of the planet overhead and the passing of the centroid of the dark spot. 

The sound speed in the photosphere sets the limit on how quickly the stellar atmosphere can react to the gravitational perturbation of the passing planet, and thus the minimum elapsed time expected between the passing of the planet and the passing of the dark spot through the sub-observer point is of order
\begin{eqnarray}
\Delta t \sim \frac{H}{c_s} \label{eqn:deltaT}
\end{eqnarray}
where $H$ is the scale height of the atmosphere, and $c_s$ is the sound speed. Using a perfect gas equation of state, we find 
\begin{eqnarray}
\Delta t \sim \sqrt{\frac{3 k_B T}{5 \mu m_H g^2}}. 
\label{eqn:Deltat}
\end{eqnarray}
The temperature ($T$) and surface gravity ($g$) at the photosphere, where the dark spot manifests, can be inferred from simulations of stellar atmospheres to calculate $\Delta t$. Robert Kurucz has calculated the pressure, temperature, density and sound speed at many depths in simulated stellar atmospheres\footnote{Models by Kurucz available on his website: \url{http://kurucz.harvard.edu/grids.html}}. We chose the nearest analogue in the grids of simulated stars to HAT-P-7 with surface temperature $T_{eff} = 6250$ K and surface gravity $\log g = 4$ \citep{pal2008}. We compute the temperature of the photosphere, which we define as the layer of the atmosphere where the optical depth $\tau \sim 2/3$. Our observed response time of $1.06 \times 10^4$ s, measured by the phase lag of the dark spot, is essentially equivalent to the Eq.\ \ref{eqn:Deltat} prediction ($\Delta t \sim 1.02 \times 10^4$ s), supporting our hypothesis that the perturbation seen in Figure \ref{phaseFoldedTransit} is due to the time-dependent gravity-darkened response of the stellar atmosphere to the passage of the planet, as predicted by J12.

\section{Summary} \label{sec:summary}
We have presented new analysis of eight quarters of \kepler photometry of HAT-P-7b. We refine the orbital parameters of the system and a smaller secondary eclipse depth than the earliest published results. These findings are consistent with other analyses of subsets of the observations that we considered. The secondary eclipse of HAT-P-7b indicates a geometric albedo of $\lesssim 0.04$, qualitatively consistent with the properties of the often compared, highly-irradiated hot Jupiter TrES-2b. We find no evidence for variations in the eclipse depth. 

The composite transit light curve of HAT-P-7b has a $\sim10$ ppm anomaly that is consistent with both the amplitude and phase lag of a planet-induced gravity-darkened spot on the stellar surface, though additional transit observations are required to confirm it. The sub-planet region of the star experiences a decrease in surface gravity, which results in a corresponding decrease in temperature and brightness that is observable in the light curve. The ongoing \kepler mission will continue to observe HAT-P-7b, and future analyses of the phase-folded transit light curve will be able to conclusively verify the astrophysical nature of the anomaly. 

\acknowledgements
B.M.M. and A.M.M. acknowledge support from the Goddard Center for Astrobiology and the NASA Astrobiology Institute, with administrative support from the University of Maryland. This paper includes data collected by the Kepler mission. Funding for the Kepler mission 
is provided by the NASA Science Mission directorate.
Some of the data presented in this paper were obtained from the Mikulski Archive for 
Space Telescopes (MAST). STScI is operated by the Association of Universities for 
Research in Astronomy, Inc., under NASA contract NAS5-26555. Support for MAST 
for non-HST data is provided by the NASA Office of Space Science via grant 
NNX09AF08G and by other grants and contracts. We thank the referee for insightful recommendations on the manuscript.

\end{document}